\documentclass{article}
\usepackage{amsmath}
\usepackage{amsfonts}
\usepackage{dsfont}
\usepackage{graphicx}
\usepackage{authblk}
\usepackage{cite}

\DeclareMathOperator{\tr}{Tr}
\DeclareMathOperator{\trop}{\hat{T}r}
\DeclareMathOperator{\tropT}{\check{T}r}

\begin{document}

\newcommand{\cc}[1]{\overline{#1}}
\newcommand{\mat}[1]{\hat{#1}}
\newcommand{\Hs}{\mathcal{H}_s}
\newcommand{\Hb}{\mathcal{H}_b}
\newcommand{\Ht}{\mathcal{H}}
\newcommand{\linop}[1]{\mathcal{L}(#1)}
\newcommand{\linHs}{\linop{\Hs}}
\newcommand{\linNN}[1]{#1^{N \times N}}
\newcommand{\linHsNN}{\linNN{\linHs}}
\newcommand{\HsN}{\mathcal{H}_s^N}
\newcommand{\linHt}{\linop{\Ht}}
\newcommand{\linHtNN}{\linNN{\linHt}}
\newcommand{\proj}[2]{|#1\rangle\langle#2|}

\newenvironment{smallpmatrix}{\left(\begin{smallmatrix}}{\end{smallmatrix}\right)}

\title{Reduced Operator Approximation for Modelling Open Quantum Systems}
\author{Agnieszka Werpachowska\footnote{a.m.werpachowska@gmail.com}}
\affil{\small Department of Physics and Astronomy, University College London, Gower Street, London WC1E 6BT, United Kingdom}

\maketitle

\begin{abstract}
We present the Reduced Operator Approximation: a simple, physically transparent and computationally efficient method of modelling open quantum systems. It employs the Heisenberg picture of the quantum dynamics, which allows us to focus on the system degrees of freedom in a natural and easy way. We describe different variants of the method, low- and high-order in the system-bath interaction operators, defining them for either general quantum harmonic oscillator baths or specialising them for independent baths with Lorentzian spectral densities. Its wide applicability is demonstrated on the examples of systems coupled to different baths (with varying system-bath interaction strength and bath memory length), and compared with the exact pseudomode and the popular quantum state diffusion approach. The method captures the decoherence of the system interacting with the bath, while conserving the total energy. Our results suggest that quantum coherence effects persist in open quantum systems for much longer times than previously thought.
\end{abstract}

\section{Introduction}

The beginning of the twentieth century launched a series of major paradigm shifts which heralded the era of modern physics. It will perhaps be surprising to the modern reader that in the advent of the revolutionary Einsteinian theory of relativity, Maxwell and Boltzmann's kinetic theory and Planck's hypothesis of quanta, the scientific world was not convinced of the fact that matter is grainy and cannot be continuously divided ad infinitum~\cite{Gora}. The seed of doubt was planted by the renowned Scottish botanist, Robert Brown, who noticed in 1827 that pollen in water suspension which he examined under his microscope displayed a very rapid, irregular, zigzag motion. The mystery of the ``vital force'' driving the Brownian motion remained unsolved for nearly 80 years, eva\-ding the pincer of conventional physics. The answer came from Einstein and Smoluchowski, who showed how the behaviour of mechanical objects is governed by the statistical properties of thermal noise, postulating the existence of molecules in the fluid and linking the diffusion strength of their motion to the friction acting on a body moving in the fluid~\cite{einstein1905,Smoluchowski1906}. The explanation of Brown's experiments, being at the same time a major diversion from the ``continuous'' Newtonian dynamics forming the core of contemporary physics, opened a whole new avenue of research into the behaviour of systems influenced with surrounding random noise, resulting in such fundamental discoveries as the fluctuation-dissipation theorem~\cite{fdt1,fdt2}. Since that time, dissipation has been shown to affect such key dynamical processes as electron transfer and transport, surface dynamics, quantum tunneling, control and nonadiabatic effects. Scientists in many disciplines, from physics through biology to social sciences, have developed increasingly powerful methods of modelling \emph{open systems}, which interact with their environment.

In many nano-scale open systems the noise influencing the dynamics arises from quantum fluctuations. Already in 1928, when Nyquist proposed the flu\-ctu\-ation-dissipation theorem~\cite{fdt1}, the quantum fluctuations were treated differently than the classical ones: the energy $k_B T$ from the classical equipartition law was replaced by the thermally averaged energy of a quantum harmonic oscillator, a distinction becoming negligible at high temperatures. This result has been followed by the development of the new branch of physics, the theory of \emph{open quantum systems}~\cite{isar,theory-open-quantum-systems,rivas2011open}. It has found applications in almost all areas of natural sciences~\cite{Benatti-notes}, such as quantum optics~\cite{Quantum-Optics}, condensed matter physics~\cite{Quantum-Dissipative-Systems}, nanotechnology~\cite{Quantum-Mesoscopic-Phenomena-and-Mesoscopic} and spintronics~\cite{Semiconductor-Spintronics-and-Quantum-Computation}, quantum information~\cite{Quantum-Computation-and-Quantum-Information}, chemistry~\cite{nmqsd,Quantum-Dynamics-of-Open-Systems} and biology~\cite{birds,Panitchayangkoon27122011,Hoyer14} or even stochastic gravity and inflationary cosmology~\cite{quantum-fields}. Furthermore, it has implications for such fundamental problems as the quantum measurement theory~\cite{Quantum-Measurement} and the emergence of classicality due to decoherence~\cite{Decoherence-and-the-Appearance}. 

A variety of methods for modelling open quantum systems exists, based on different mathematical techniques and applicable to different physical regimes~\cite{metoda1,metoda2,metoda3,metoda4,metoda5,strunz,pseudomode,pseudomode2}. The most popular ones employ Born and Markov approximations, leading to the general Markovian master equation in the Kossa\-kowski--Lindblad form~\cite{Kossakowski72,Gorini76,Lindblad76} or the Redfield equation~\cite{Redfield65}. To describe systems beyond the regime of small coupling with environment and short environment correlation times, more complex methods have been proposed, from formally exact generalised master equations and their numerical approximations~\cite{Esposito03,Budini05,Ishizaki09,Moodley09} to stochastic methods~\cite{metoda5,Suess14} and numerical path integral techniques~\cite{Muhlbacher08,Nalbach11}. However, they remain generally unsuitable or difficult to apply to large systems, unrestricted system-environment coupling strengths and arbitrary environment spectral densities---the properties necessary to quantitatively analyse many physical situations encountered in the fields mentioned in the previous paragraph~\cite{PhysRevD.51.1577,Bratsun11102005,Roszak}, while the approximations they make sometimes lack physical transparency and controllability. In this paper we propose a new method, reduced operator approximation (ROA), which describes finite-dimensional quantum systems up to a moderately large size (ca 50 basis states on a standard PC), interacting with non-Markovian quantum harmonic oscillator baths (from single modes to wide continuous spectra) with a wide range of coupling strengths, while having moderate computational requirements. It uses the Heisenberg picture, which makes it particularly easy to focus the attention on the system degrees of freedom while preserving the decoherence effects due to the coupling to the bath, and conserving the total energy. In this non-Markovian setting, our method goes beyond the Kossakowski--Lindblad formalism, as the generator of the evolution of system operators in the Heisenberg picture depends on their history.

In the following section we will remind shortly the theoretical background of our work and lay out the employed formalism (Secs\,\ref{sec:open-qs} and~\ref{sec:dynamics-Heisenberg}). Next we will present the derivation of ROA (Sec.\,\ref{sec:ROA}) and propose its two variants: low and high-order in the system-bath interaction operators. They will be optimised for typical cases of continuous and Lorentzian baths in Sec.\,\ref{sec:baths}. In Sec.\,\ref{sec:examples} we will apply our method to open quantum systems in different system-bath coupling strength and bath memory regimes, and compare it to other popular modelling techniques, such as the pseudomode method and the quantum state diffusion. Section\,\ref{sec:summary} contains a short summary of our work.

\section{Theoretical approach}
\label{sec:approach}

Most generally, an open quantum system is a subsystem of a larger, interacting quantum system, e.g.~one of the photons in an EPR pair, an atom in a resonant cavity, a quantum dot interacting with phonons in the crystal or any real object ``becoming classical'' through scattering of a vast number of air molecules and photons on it. We consider the case of a finite-dimensional quantum system coupled to an infinite-dimensional quantum bath, composed of a possibly infinite number of modes. In such an asymmetrical setup it is natural to ignore the details of the bath dynamics and focus on the dynamics of the reduced density matrix of the system. In this chapter we derive it using the proposed ROA approach.

\subsection{Open quantum system}
\label{sec:open-qs}
\setcounter{equation}{0}

We consider a quantum system represented in an $N$-dimensional Hilbert space $\Hs$ spanned by basis states $\{ |n\rangle \}$ and associated with the space of linear operators on $\Hs$, denoted by $\linHs$. Its internal dynamics is described by the Hamiltonian $H_s \in \linHs$,
\[
H_s = \sum_{m,n=1}^N V_{mn} t_{mn} \,,
\]
where $\linHs \ni t_{mn} := \proj{m}{n}$ are transition operators between the states $|n\rangle$ and $|m\rangle$ and $V_{mn} \in \mathbb{C}$ are the interstate couplings, $\cc{V_{mn}} = V_{nm}$. In a more concise notation, $H_s$ is a trace of a matrix product in $\linHsNN$, the space of $N \times N$ matrices whose elements are operators from $\linHs$:
\[
H_s = \trop \mat{V}^T \mat{t} := \sum_{n=1}^N (\mat{V}^T \mat{t})_{nn} \,.
\]
In the equation above, $\mat{t} \in \linHsNN$ is an $N \times N$ matrix of system operators, $(\mat{t})_{mn} := t_{mn}$, and $\mat{V} \in \mathbb{C}^{N \times N}$. Since $z \in \mathbb{C}$ can be embedded in $\linHs$ as $z \mathbb{I}_s$ (where $\mathbb{I}_s$ is the identity operator on $\Hs$), the matrix product between $\mathbb{C}^{N\times N}$ and $\linHsNN$ can be defined, with values in $\linHsNN$. $\trop$ denotes the trace in $\linHsNN$,
\[
\linHs \ni \trop \mat{O} := \sum_{n=1}^N O_{nn} \ , \quad \mat{O} \in \linHsNN \,.
\]
By a natural extension of the scalar product from $\Hs$,
\[
\langle \Psi | \Phi \rangle_{\HsN} := \sum_{n=1}^N \langle \Psi_n | \Phi_n \rangle_{\Hs}\,,
\]
and 
$\HsN := (\Hs)^N$ is a Hilbert space associated with the space of linear operators $\linop{\HsN} \equiv \linHsNN$.
Hermitian conjugation in $\linHsNN$ can be thus defined as $(\mat{O}^\dagger)_{mn} := O_{nm}^\dagger$. In the case of $\mat{t}$, this leads to $\mat{t}^\dagger = \mat{t}$, since $t_{mn}^\dagger = t_{mn}$.

The system is coupled to a quantum bath composed of a collection of $K$ independent harmonic oscillators described by the Hamiltonian acting on an infinite-dimensional Hilbert space $\Hb$,
\[
\linop{\Hb} \ni H_b = \sum_{k=1}^K \omega_k a_k^\dagger a_k \,,
\]
where $a_k$ is the annihilation operator of the $k$-th mode ($\hbar \equiv 1$). The coupling between the system and the bath is described by the operator
\[
\linop{\Ht} \ni H_i = \sum_{k=1}^K \left(\trop \mat{g}_k \mat{t}\right) a_k^\dagger + \mathrm{h.c.}\,,
\]
where $\Ht = \Hs \otimes \Hb$ and $\{\mat{g}_k\}$ are $N\times N$ diagonal matrices describing the coupling of the $k$-th bath mode with the system, $(\mat{g}_k)_{mn} = \delta_{mn} g_{kn}$.

The fact that each $\mat{g}_k$ matrix is diagonal means that the bath does not induce transitions between system basis states. However, the matrix notation allows for an easy generalisation of the model to include such bath-induced transitions. 

The total Hamiltonian generating the evolution of the system and the bath in the Schr\"odinger picture is given by
\begin{equation}
H = H_s + H_b + H_i =
\trop \mat{V}^T \mat{t} + \sum_{k=1}^K \omega_k a_k^\dagger a_k + \trop \sum_{k=1}^K \left( a_k^\dagger \mat{g}_k + a_k \mat{g}_k^\dagger \right) \mat{t} \,.
\label{Hamiltonian}
\end{equation}
In the traditional ``index'' notation
\[
H = \sum_{m,n=1}^N V_{mn} t_{mn} + \sum_{k=1}^K \omega_k a_k^\dagger a_k + \sum_{m=1}^N \sum_{k=1}^K ( \cc{g}_{km} a_k + g_{km} a_k^\dagger ) t_{mm}\,.
\]

The reduced density matrix of the system is defined as 
\[
\rho_s(t) := {\tr}_b \rho(t)\,,
\]
where $\rho(t)$ is the density matrix of the system and the bath as a whole and the trace ${\tr}_b : \linHt \mapsto \linHs$ is performed over the bath degrees of freedom only. 
In the $|n\rangle$ basis,
\begin{equation}\label{rho-Schr}
(\rho_s(t))_{mn} = \tr [\rho(t) t_{mn}] \,,
\end{equation}
where the trace is over both system and bath degrees of freedom. The main task of the presented method is obtaining $\rho_s(t)$ without calculating $\rho(t)$.

\subsection{Dynamics in the Heisenberg picture}
\label{sec:dynamics-Heisenberg}

In the Heisenberg picture the wavefunction is time-independent, $\Psi \equiv \Psi(0)$ (hence, the density matrix is time-independent as well), while an observable $O$ (time-independent in the Schr\"{o}dinger picture) satisfies
\begin{equation}\label{Heisenberg}
\frac{d}{dt} O(t) = i [ H , O(t) ] \,.
\end{equation}
where $O(t) := e^{iHt} O e^{-iHt}$. From the last definition follows $[O_1(t), O_2(t)] = [O_1, O_2](t)$. 

We assume that at time $t = 0$ the system and the bath---represented by their initial reduced density matrices $\rho_s$ and $\rho_b$, respectively---are uncorrelated. Hence, the total density matrix $\rho$ equals $\rho_s\otimes\rho_b$ and \eqref{rho-Schr} becomes
\begin{equation}\label{rho-Heis}
(\rho_s(t))_{mn} = \tr [t_{mn}(t) \rho] = \tr [t_{mn}(t) \rho_s\otimes\rho_b] = {\tr}_s [\rho_s {\tr}_b \rho_b t_{mn}(t)] \,.
\end{equation}

Let $t_{mn}(t), a_k(t) \in \linHt$ denote the Heisenberg-picture counterparts of $t_{mn}$ and $a_k$, with $t_{mn}(0) := t_{mn} \otimes \mathbb{I}_b$ (where $\mathbb{I}_b$ is the identity operator on $\Hb$) and $a_k(0) := \mathbb{I}_s \otimes a_k$. The time-dependent operators act on the complete Hilbert space $\Ht$, reflecting the fact that the interaction couples the system and the bath. Analogously to the Schr\"{o}dinger picture, we define $\linHtNN \ni (\mat{t}(t))_{mn} := t_{mn}(t)$. We also define the trace $\tropT : \linHtNN \mapsto \linHt$ as
\[
\tropT \mat{O} := \sum_{n=1}^N O_{nn}\ ,\ \ \mat{O} \in \linHtNN\,.
\]

In order to derive equations of motions for $t_{mn}(t)$ and $a_k(t)$ using \eqref{Heisenberg} we calculate, for any $\mat{A} \in \mathbb{C}^{N\times N}$ (which can be embedded in $\linHt^{N \times N}$ by embedding each element $A_{mn}$ in $\linHt$),
\begin{equation}
[\tropT (\mat{A} \mat{t}(t)), t_{mn}(t) ] = \sum_{m',n'=1}^N A_{m'n'} [ t_{n'm'}(t), t_{mn}(t) ] 
= [\mat{A},\mat{t}(t)]_{mn} \,,
\end{equation}
where we have used the identity (valid also in the Schr\"{o}dinger picture)
\begin{equation}\label{tt-elements}
t_{mn}(t) t_{m'n'}(t) \equiv t_{mn'}(t) \delta_{nm'} \,.
\end{equation}
In the more compact notation, $[ \tropT (\mat{A} \mat{t}(t)), \mat{t}(t) ] = [\mat{A}, \mat{t}]$.

The above identities can be applied to \eqref{Heisenberg} to obtain the evolution of $\mat{t}$ and $\mat{a}$ operators generated by the Hamiltonian~\eqref{Hamiltonian}. For the system we obtain
\begin{equation}\label{tmn-dyn}
\dot{ \mat{t}}(t) 
= i [ \mat{V}^T, \mat{t}(t) ] + i \sum_{k=1}^K \left( [ \mat{g}_k, \mat{s}_k^\dagger(t) ] + [ \mat{g}_k^\dagger, \mat{s}_k] \right) ,
\end{equation}
where $\mat{s}_k(t) := \mat{t}(t) a_k(t)$ are system-bath interaction operators and, since $\mat{t}^\dagger(t) = \mat{t}(t)$ and bath operators commute with system operators, $\mat{s}^\dagger_k(t) = \mat{t}(t) a_k^\dagger(t)$. For the bath, using the canonical commutation relations for bosonic cre\-a\-tion/anni\-hi\-lation operators, we obtain
\begin{equation}\label{ak-dyn}
\dot{a}_k(t) = -i \omega_k a_k(t) - i \tropT \mat{g}_{k} \mat{t}(t)  \,.
\end{equation}

In the index notation, the above equations have the following form (assuming $(\mat{g}_k)_{mn} = \delta_{mn} g_{kn}$):
\begin{equation}
\begin{split}
\label{tmn-dyn-idx}
&\frac{d}{dt} t_{mn}(t) = i \sum_{m'=1}^N \left( V_{m'm} t_{m'n}(t) - V_{nm'} t_{mm'}(t) \right)\\
&\qquad+ i t_{mn}(t) \sum_{k=1}^K ( g_{km} - g_{kn} ) a_k^\dagger(t) + i t_{mn}(t) \sum_{k=1}^K ( \cc{g_{km}} - \cc{g_{kn}} ) a_k(t)
\end{split}
\end{equation}
and
\[
\frac{d}{dt} a_k(t) = -i \omega_k a_k(t) -i \sum_{m=1}^N g_{km} t_{mm}(t)\,.
\]

\subsection{Reduced Operator Approximation}
\label{sec:ROA}

\subsubsection{General description}
\label{sec:genver}

The aim of the presented method is to model the evolution of the system, including its decoherence caused by the interaction with the bath. The information about this process is contained in the reduced density matrix of the system $\rho_s(t)$. As demonstrated in previous sections, see \eqref{rho-Schr} and \eqref{rho-Heis}, it can be obtained from the mean values of system operators $\mat{t}(t)$. Thus, to calculate $\rho_s(t)$ in the Heisenberg picture, one has to evolve $\mat{t}(t)$ in time. Since the evolution equation for $\mat{t}(t)$~\eqref{tmn-dyn} involves the bath operators $a_k(t)$, due to the system-bath interaction, it is necessary to evolve $a_k(t)$ as well. However, a numerical description of both types of operators in the total system and bath basis is impossible, as $\mathcal{H}_b$ is infinite-dimensional.

According to \eqref{rho-Heis}, given the initial system state $\rho_s$ we only need to know the partial expectation values of system operators, ${\tr}_b \rho_b t_{mn}(t)$, to obtain $\rho_s(t)$. The corresponding partial expectations of the bath operators, ${\tr}_b \rho_b a_k(t)$, contain part of the information on how interaction correlated the bath and the system---if there was no such correlation, ${\tr}_b \rho_b a_k(t)$ would be proportional to an identity operator in $\mathcal{H}_s$ (recall that $a_k(0) = \mathbb{I}_s \otimes a_k$). Thus, even after averaging over the bath degrees of freedom, we can at least approximately capture the system-bath correlations arising from the interaction terms in \eqref{tmn-dyn} and~\eqref{ak-dyn}. This observation forms the basis of ROA.

We represent both system and bath operators by $N \times N$ complex matrices in the system state basis (hence, $\mat{t}$ is represented by an $N \times N$ matrix of $N \times N$ complex matrices). Let $M[O(t)] \in \linHs$ denote this \emph{reduced representation} of an operator $O(t) \in \linHt$, defined as 
\[
M[O(t)] := {\tr}_b \rho_b O(t) \,.
\]
The mapping $M[O(t)]$ preserves, at least partially, the information about how the operator $O(t)$ acts on the system degrees of freedom in the presence of interaction with the bath. Since the system-bath interaction depends on the initial state of the bath $\rho_b$, so does $M[O(t)]$, the effect particularly strong in non-Markovian systems. To simplify the derivation of the ROA evolution equations, we assume that at $t = 0$ the bath was in its ground state, $\rho_b = \proj{\Omega_b}{\Omega_b}$, where $\Omega_b = \bigotimes_{k=1}^K |0\rangle_k$ and $|0\rangle_k$ is the ground state of the $k$-th bath mode.

As can be seen from \eqref{rho-Heis}, the mapping $M[\cdot]$ conserves the expectation values:
\[
\langle O(t) \rangle = {\tr} \rho O(t) = {\tr}_s [\rho_s {\tr}_b \rho_b O(t)] = {\tr}_s [\rho_s M[O(t)]] = \langle M[O(t)] \rangle\,.
\]
It is easily extended elementwise to $M[\mat{O}(t)] \in \linHsNN$ for $\mat{O}(t) \in \linHtNN$ and, from the definition, $M[O(t)^\dagger] = M[O(t)]^\dagger$.

The evolution equations for the reduced representations of system and bath operators are
\begin{equation}
\label{Mak-dyn}
\frac{d}{dt} M[a_k(t)] = -i \omega_k M[a_k(t)] - i {\tr} \mat{g}_{k} M[\mat{t}(t)]
\end{equation}
and
\\
\begin{equation}
\label{tmn-dyn-MFAHO}
\frac{d}{dt} M[\mat{t}(t)] = i [ \mat{V}^T, M[\mat{t}(t)] ] + i \sum_{k=1}^K \left( [ \mat{g}_k, M[\mat{s}_k^\dagger(t)] ] + [ \mat{g}_k^\dagger, M[\mat{s}_k(t)] ] \right).
\end{equation}

Since the system and the bath are correlated, $M[ \mat{s}_{k}(t) ] \neq M [ \mat{t}(t) ] M [ a_k(t) ]$, which means that the above evolution equations are not complete. The simplest way to complete them is to approximate $M[\mat{s}_k(t)]$ by the product of $M[\mat{t}(t)]$ and $M[a_k(t)]$, which means neglecting higher-order correlations between the system and the bath introduced by their interaction. However, again due to the system-bath coupling, $M[\mat{t}(t)] M[a_k(t)] \neq M[a_k(t)] M[\mat{t}(t)]$, even though $[\mat{t}(t), a_k(t)] = 0$. Thus, we need to specify an ordering of the multiplied reduced operators. We use the approximations of the form
\[
\begin{split}
&M[\mat{s}_k(t)] \approx \theta_l M[\mat{t}(t)] M[a_k(t)] + (1-\theta_l) M[a_k(t)] M[\mat{t}(t)] \,,\\
&M[\mat{s}^\dagger_k(t)] \approx \theta_l M[a_k(t)]^\dagger M[\mat{t}(t)] + (1-\theta_l) M[\mat{t}(t)] M[a_k(t)]^\dagger ,
\end{split}
\]
for $\theta_l \in [0, 1]$. Based on our numerical experiments, which have shown that simulations diverge for $\theta_l \neq \frac{1}{2}$, we put $\theta_l = \frac{1}{2}$. Appendix\,A contains an examination of this problem for a simplified case, supporting this choice. Furthermore, the symmetrization of the product of $M[a_k(t)]$ and $M[\mat t(t)]$ is consistent with the fact that $a_k(t)$ and $\mat t(t)$ always commute. Thus, the final form of the evolution equation of the system operators in the reduced representation is
\begin{equation}
\begin{split}
\label{tmn-dyn-MFA}
\frac{d}{dt} M[\mat{t}(t)] &= i [ \mat{V}^T, M[\mat{t}(t)] ] 
 + \frac{i}{2} \sum_{k=1}^K [ \mat{g}_k, \{ M [ \mat{t}(t) ], M [ a_k^\dagger(t) ] \}]
\\&+ \frac{i}{2} \sum_{k=1}^K [ \mat{g}_k^\dagger, \{ M [ \mat{t}(t) ], M [ a_k(t) ] \}] \,,
\end{split}
\end{equation}
where $\{\cdot,\cdot\}$ denotes the anti-commutator. Equations~\eqref{Mak-dyn} and~\eqref{tmn-dyn-MFA} employ reduced representations which are linear in the system or bath operators. Hence, we will refer to them as the \emph{lower-order} ROA. It is important to note that this form does not neglect the correlations between the system and the bath, because the bath operators are represented by their matrices in the system basis, $M[a_k(t)]$.

Additional information about the system-bath correlations is provided by the $M[\mat{s}_k(t)]$ matrix. Hence, it may be beneficial to evolve it separately in addition to $M[a_k(t)]$ and $M[\mat{t}(t)]$. For this purpose, we first derive the evolution equation for $\mat{s}_k(t)$, using \eqref{tmn-dyn} and~\eqref{ak-dyn},
\[
\begin{split}
\frac{d}{dt} \mat{s}_k(t) &= \mat{t}(t) \frac{d}{dt} a_k(t) + \left( \frac{d}{dt} \mat{t}(t) \right) a_k(t) 
= i [ \mat{V}^T, \mat{s}_k(t) ] + i \sum_{k'=1}^K  [ \mat{g}^\dagger_{k'}, \mat{s}_{k}(t) ] a_{k'}(t)
 \\&+ i \sum_{k'=1}^K a_{k'}^\dagger(t) [ \mat{g}_{k'}, \mat{s}_{k}(t) ] - i \omega_k \mat{s}_{k}(t) - i \mat{t}(t) \mat{g}_k \,,
\end{split}
\]
where we have used the fact that, due to the associativity of the operator product, $\mat{s}_{k'}(t) a_{k}(t) = \mat{t}(t) a_{k'}(t) a_k(t) = a_{k'}(t) \mat{s}_{k}(t)$ and $\mat{s}_{k'}^\dagger(t) a_{k}(t) = a_{k'}^\dagger(t) \mat{s}_{k}(t)$. To derive the evolution equation for $M[\mat{s}_k(t)]$, we have to solve a similar ordering problem as in the lower-order method:
\[
\begin{split}
M[\mat{s}_{k}(t) a_{k'}(t)] &\approx \theta_h M[\mat{s}_{k}(t)] M[a_{k'}(t)] + (1 - \theta_h) M[a_{k'}(t)] M[\mat{s}_{k}(t)]\,,\\
M[a_{k'}^\dagger(t) \mat{s}_{k}(t)] &\approx \theta_h M[a_{k'}^\dagger(t)] M[\mat{s}_{k}(t)] + (1 - \theta_h) M[\mat{s}_{k}(t)] M[a_{k'}^\dagger(t)] \\
&\qquad- (1 - \theta_h) \delta_{k,k'} M[\mat{t}(t)]\,,
\end{split}
\]
where the last term arises from the fact that $a_{k'}^\dagger(t)$ and $\mat{s}_{k}(t)$ do not commute, unlike $a_k(t)$ and $\mat t(t)$ from the lower-order case. For the bath in the ground state at $t = 0$ we must have $M[a_{k'}^\dagger(0) \mat{s}_{k}(0)] = 0$. It is easy to see that the only way this can be satisfied is by choosing $\theta_h = 1$, as otherwise $M[a_{k'}^\dagger(0) \mat{s}_{k}(0)] = - (1 - \theta_h) \delta_{k,k'} M[\mat{t}(t)] \neq 0$. This choice is supported by numerical tests, which yield divergences when $\theta_h \neq 1$ is used. For consistency, we apply the same choice of $\theta_h$ to the approximation of $M[\mat{s}_{k}(t) a_{k'}(t)]$. Another problem is that $M[\mat{s}_{k}(t) a_{k'}(t)]$ can be approximated by either $M[\mat{s}_k(t)] M[a_{k'}(t)]$ or $M[\mat{s}_{k'}(t)] M[a_{k}(t)]$. Similarly, we can choose between $M[a_{k'}^\dagger(t)] M[\mat{s}_{k}(t)]$ and $M[\mat{s}^\dagger_{k'}(t)] M[a_{k}(t)]$ for $M[a_{k'}^\dagger(t) \mat{s}_k(t)]$. To exploit fully the information about the system-bath correlations contained in $M[\mat{s}_{k'}(t)]$ matrices, we use an equally weighted average of the two approximations. In this way we obtain the evolution equation for the reduced representation of interaction operator
\[
\begin{split}
\frac{d}{dt} M[\mat{s}_k(t)] &=  -i \omega_k M[\mat{s}_{k}(t)] - i M[\mat{t}(t)] \mat{g}_k + i [ \mat{V}^T, M[\mat{s}_k(t)] ]\\
&+ \frac{i}{2} \sum_{k'=1}^K \left( [ \mat{g}_{k'}, M[\mat{s}_{k'}^\dagger(t)] ]
+ [ \mat{g}_{k'}^\dagger, M[\mat{s}_{k'}(t)] ] \right) M[a_k(t)]\\
&+ \frac{i}{2} \sum_{k'=1}^K M[a_{k'}^\dagger(t)] [\mat{g}_{k'}, M[\mat{s}_{k}(t)] ] 
+ \frac{i}{2} \sum_{k'=1}^K  [ \mat{g}_{k'}^\dagger, M[\mat{s}_k(t)] ] M[a_{k'}(t)]\,.
\end{split}
\]
Together with \eqref{Mak-dyn} and~\eqref{tmn-dyn-MFAHO} it defines the \emph{higher-order} ROA.

\subsubsection{Total Hamiltonian and energy conservation}

In the lower-order version of our method, the total Hamiltonian is
\[
\begin{split}
M[H] &= \trop \mat{V}^T M[\mat{t}(t)] + \sum_{k=1}^K \omega_k M[a_k^\dagger(t)] M[a_k(t)] 
\\&+ \trop \sum_{k=1}^K \left\{ \left( M[a_k^\dagger(t)] \mat{g}_k + M[a_k(t)] \mat{g}_k^\dagger \right), M[\mat{t}(t)] \right\}.
\end{split}
\]
In the higher-order version, the last term acquires the form 
\[\trop \sum_{k=1}^K \left( M[\mat{s}_k^\dagger(t)] \mat{g}_k + M[\mat{s}_k(t)] \mat{g}_k^\dagger \right)\,.
\]

A tedious but straightforward calculation shows that in both cases, the evolution equations from Sec.~\ref{sec:genver} conserve $M[H]$. Thus, our method conserves the total energy. 

\subsubsection{Reduced system density matrix}
\label{sec:reduced-system-density-matrix}

From the evolution equations~\eqref{tmn-dyn-MFAHO} or~\eqref{tmn-dyn-MFA} we instantly see that, since the trace of every commutator is zero,
\[
\frac{d}{dt} \left( {\tr} M[\mat{t}(t)] \equiv \sum_{m=1}^N M[t_{mm}(t)] \right) = 0\,.
\]
Inserting $M[\mat{t}(t)]$ instead of $\mat{t}(t)$ in \eqref{rho-Heis}, we obtain a trace-conserving expression for the reduced density matrix of the system, $(\rho_s(t))_{mn} = {\tr}_s \rho_s M[t_{mn}(t)]$. However, if we use it to calculate $\rho_s(t)$, its positivity is not guaranteed. To fix this problem, we make use of the identity~\eqref{tt-elements} to derive 
$\mat{t}(t) \mat{t}(t) \equiv N \mat{t}(t)$ and replace \eqref{rho-Heis} with a different approximation
\begin{equation}
\label{rho-Heis-positive}
(\rho_s(t))_{mn} = \frac{1}{N} {\tr}_s \rho_s M[t_{mn}(t)]^2 \,.
\end{equation}
Since $t_{mm'}(t) t_{m'n}(t) = t_{mm'}(t) (t_{nm'}(t))^\dagger$ (and the same for the reduced representations), the above formula guarantees that $\rho_s(t)$ is positive-semidefinite. In contrast, the trace of density matrix~\eqref{rho-Heis-positive} is not conserved, because $M[\mat{t}(t)]^2 \neq M[\mat{t}(t)^2]$. Thus, we normalise the density matrix to obtain
\[
(\rho_s(t))_{mn} = {\tr}_s \rho_s M[t_{mn}(t)]^2/\sum_{n'=1}^N{\tr}_s \rho_s M[t_{n'n'}(t)]^2\,.
\]

\subsection{Baths with continuous spectral densities}
\label{sec:baths}

For any type of bath we can define the spectral density $\mat{J}(\omega) = \sum_k \mat{g}_k^\dagger \mat{g}_k \delta(\omega - \omega_k)$. In the limit of an infinite number of modes, $\mat{J}(\omega)$ can be a continuous function. One way to handle this situation is to discretise $\mat{J}(\omega)$ into a finite, but large number of modes. Assuming a constant mode frequency spacing $\Delta \omega$, we define a coupling constant for $\omega = k \Delta\omega$ to be
\begin{equation}
\mat{g}_{k} = \sqrt{ \Delta \omega \mat{J}(k \Delta \omega) } \,.
\label{gk}
\end{equation}
Taking the square root ensures proper normalisation as $\Delta\omega \to 0$. For a spectral density being a single Lorentzian peak, the method converges quite well already for $K=100$ modes per system basis state and $\Delta \omega \approx \gamma/100$, where $\gamma$ is the half-width at half-maximum of the peak.

\subsubsection{Independent baths with continuous spectral densities}
\label{sec:independent-baths}

When each system basis state is coupled to its own independent bath with a continuous spectral density, it is more convenient to describe these baths in terms of collective mode excitations caused by the coupling with the system.

Let us consider bath operators multiplied by their phase factors, $e^{i\omega_k t} a_k(t)$, with dynamics given by $\frac{d}{dt} (e^{i\omega_k t} a_k(t)) = - i \tropT \mat{g}_{k} \mat{t}(t) e^{i\omega_k t}$.
Employing the fact that taking the trace and integration over time commute, we obtain
\[
a_k(t) = e^{-i\omega_k t} a_k(0) - i \int_0^t ds e^{-i\omega_k (t-s)} \tropT \mat{g}_{k} \mat{t}(s)\,.
\]
We insert it into the definition of system-bath interaction operator $\mat{s}_k(t)$, 
\[
\mat{s}_k(t) = e^{-i\omega_k t} \mat{t}(t) a_k(0) - i \mat{t}(t) \int_0^t ds e^{-i\omega_k (t-s)} \tropT \mat{g}_{k} \mat{t}(s) = e^{-i\omega_k t} \mat{t}(t) a_k(0) + \mat{u}_k(t)\,,
\]
and then insert the obtained formula into \eqref{tmn-dyn}:
\[
\begin{split}
\dot{ \mat{t}}(t) &= i [ \mat{V}^T, \mat{t}(t) ] + i \sum_{k=1}^K \left( [ \mat{g}_k, \mat{u}_k^\dagger(t) ] + [ \mat{g}_k^\dagger, \mat{u}_k(t)] \right) 
\\&+ i \sum_{k=1}^K \left( e^{i\omega_k t} a_k^\dagger(0) [ \mat{g}_k, \mat{t}(t) ] + e^{-i\omega_k t} [ \mat{g}_k^\dagger, \mat{t}(t)] a_k(0) \right).
\end{split}
\]

Under the above assumptions, each bath mode $k$ is coupled to exactly one basis state $n_k$, i.e.~$(\mat{g}_k)_{mn} = g_{kn_k} \delta_{m,n_k} \delta_{n,n_k}$.
Hence,
\[
\begin{split}
\frac{d}{dt} t_{mn}(t) =& i \sum_{m'=1}^N (V_{m'm} t_{m'n}(t) - V_{nm'} t_{mm'}(t))
\\&+ t_{mn}(t)\sum_{k=1}^K \left[  \int_0^t e^{-i\omega_k (t-s)} (|g_{km}|^2 t_{mm}(s) - |g_{kn}|^2 t_{nn}(s)) ds - \mathrm{h.c.} \right] 
\\&+ t_{mn}(t) \sum_{k=1}^K \left[ (\cc{g_{km}} - \cc{g_{kn}}) e^{-i\omega_k t} a_k(0) + \mathrm{h.c.} \right]\,,
\end{split}
\]
where we have used the fact that for independent densities, $g_{kn} \cc{g_{km}} \neq 0$ only for $n = m$.
In the limit of infinite number of modes, using \eqref{gk} we obtain
\[
\lim_{K \to \infty} \sum_{k=1}^K e^{-i\omega_k (t-s)} |g_{km}|^2 = \alpha_m(t - s)\,,
\]
where $\alpha_m(\tau) := \int d\omega J_m(\omega) e^{-i\omega\tau}$ is the bath correlation function. In this way we derive a closed system of differential-integral equations for $t_{mn}(t)$,
\begin{equation}
\begin{split}
\label{tmn-diffintegr}
\frac{d}{dt} t_{mn}(t) =& i \sum_{m'=1}^N (V_{m'm} t_{m'n}(t) - V_{nm'} t_{mm'}(t))
\\&+ t_{mn}(t) \left[ \sqrt{\kappa_m} \tilde{a}_m(t)
- \sqrt{\kappa_n} \tilde{a}_n(t) - \mathrm{h.c.} \right]
\end{split}
\end{equation}
where 
\[
\begin{split}
&\tilde{a}_m(t) := \frac{i}{ \sqrt{\kappa_m} }\lim_{K \to \infty} \sum_{k=1}^K \cc{g_{km}} a_k(t) \\
&= \frac{i}{\sqrt{\kappa_m}} \lim_{K \to \infty} \sum_{k=1}^K \cc{g_{km}} e^{-i\omega_k t} a_k(0)
+ \int_0^t \frac{\alpha_m(t-s)}{\sqrt{\kappa_m}} t_{mm}(s) ds
\end{split}
\]
and
\[
\kappa_m := \lim_{K \to \infty} \sum_{k=1}^K |g_{km}|^2 = \int_{-\infty}^\infty J_m(\omega) d\omega \,.
\]

Operators $\tilde{a}_m^\dagger(t)$ and $\tilde{a}_m(t)$ satisfy canonical commutation relations,\linebreak $[\tilde{a}_m(t), \tilde{a}_n(t) ] = 0$ and $[\tilde{a}_m^\dagger(t), \tilde{a}_n(t) ] = - \frac{\delta_{mn}}{\sqrt{\kappa_m \kappa_n}} \lim_{K \to \infty} \sum_{k=1}^K \cc{g_{kn}} g_{km}$ $= - \delta_{mn}$. They are \emph{pseudomode} creation and annihilation operators, creating or destroying collective excitations in a single bath~\cite{pseudomode}. Their dynamics is described by the equation
\begin{equation}
\begin{split}
\label{am-evol}
\frac{d}{dt} \tilde{a}_m(t) =& \lim_{K \to \infty} \sum_{k=1}^K \cc{g_{km}} \frac{\omega_k e^{-i\omega_k t} a_k(0)}{\sqrt{\kappa_m}} + \frac{\alpha_m(0)}{\sqrt{\kappa_m}} t_{mm}(t)
\\&+ \kappa_m^{-1/2} \int_0^t t_{mm}(s) \left( \frac{d}{dt} \alpha_m(t - s) \right) ds \,.
\end{split}
\end{equation}

Using the proposed method we have reduced significantly the number of bath operators, from $K$ to $N$ (for independent baths $K \ge N$, while in many cases $K \gg N$). However, numerical simulation of the differential-integral equation~\eqref{am-evol} for the evolution of the reduced representation of $\tilde{a}_m(t)$ is difficult. In the next section we will show that for a particular form of the spectral density function $J_m(\omega)$ one can get rid of the explicit time integration at the cost of a moderate increase of the number of simulated bath operators.

\subsubsection{Lorentzian spectral densities}
\label{sec:Lorentzian-spectral-densities}

Continuous spectral densities composed of Lorentzian peaks,
\begin{equation}
\label{JLor}
J_n(\omega) = \sum_j \frac{\Gamma_{nj} }{\pi} \frac{\gamma_{nj}}{(\omega - \omega_{nj})^2 + \gamma_{nj}^2} \,,
\end{equation}
are especially popular due to their analytical tractability. In this section, we will optimise our method for this type of the system-bath coupling. The corresponding correlation function is $\alpha_m(\tau) = \sum_j \Gamma_{mj} e^{-i\omega_{mj} \tau - \gamma_{mj} | \tau | }$. 
Hence, for $t - s > 0$, $\frac{d}{dt} \alpha_m(t - s) = -\sum_j \Gamma_{mj} (i\omega_{mj} + \gamma_{mj}) e^{-i\omega_{mj} (t-s) - \gamma_{mj} ( t -s ) }$.

A continuous spectral density of the form~\eqref{JLor} is constructed from an infinite number of independent harmonic oscillator modes, with different modes contributing to each Lorentzian peak. To derive the evolution equation for pseudomode bath operators $\tilde{a}_m(t)$ we express them as sums of
\[
\tilde{a}_{mj}(t) = \frac{i}{\sqrt{\Gamma_{mj}}} \lim_{K\to\infty} \sum_{k \in P^K_j} \cc{g_{km}} a_k(t)\,,
\]
where $P^K_j \subset [1, \dots, K]$ is the set of indices of modes building the $j$-th peak, $\cup_j P^K_j = [1,\dots,K]$ and $P_j^K \cap P_{j'}^K = \delta_{jj'} P_j^K$. This leads directly to\linebreak $[ t_{mn}(t), \tilde{a}_{m'j'}(t) ]$ $=$ $0$, $[ \tilde{a}_{mj}(t), \tilde{a}_{nj'}(t) ]$ $=$ $0$ and $[ \tilde{a}^\dagger_{mj}(t), \tilde{a}_{nj'}(t) ] = - \delta_{mn} \delta_{jj'}$. Thus, $\tilde{a}^\dagger_{mj}(t)$ and $\tilde{a}_{mj}(t)$ are pseudomode creation and annihilation operators corresponding to individual Lorentzian peaks in bath spectral densities.

Comparing $\tilde{a}_{mj}(t)$ with $\tilde{a}_m(t)$ leads to 
\[
\tilde{a}_{mj}(t)= \tilde{a}^{(0)}_{mj}(t) + \sqrt{\Gamma_{mj}}  \int_0^t e^{(i\omega_{mj} + \gamma_{mj})(s-t)} t_{mm}(s) ds \,,
\]
where $\tilde{a}^{(0)}_{mj}(t) := \frac{i}{\sqrt{\Gamma_{mj}}} \lim_{K\to\infty} \sum_{k \in P^K_j} \cc{g_{km}} e^{-i\omega_k t} a_k(0)$ and $M[\tilde{a}^{(0)}_{mj}(t)] = 0$. Differentiating over $t$ gives
\[
\begin{split}
\frac{d}{dt} \tilde{a}_{mj}(t) =& \lim_{K\to\infty} \sum_{k \in P^K_j} \frac{\cc{g_{km}} \omega_k e^{-i\omega_k t}}{\sqrt{\Gamma_{mj}}} a_k(0)  \\
&- (i \omega_{mj} + \gamma_{mj}) \left[ \tilde{a}_{mj}(t) - \tilde{a}^{(0)}_{mj}(t) \right]
+ \sqrt{\Gamma_{mj}} t_{mm}(t)\,,
\end{split}
\]
with the initial condition $\tilde{a}_{mj}(0) = \tilde{a}^{(0)}_{mj}(0)$. By splitting $\tilde{a}_m(t)$ into a sum of $\tilde{a}_{mj}(t)$, we have simplified the differential-integral evolution equation~\eqref{am-evol}. In the reduced representation,
\begin{equation}
\label{atilde-reduced}
\frac{d}{dt} M[\tilde{a}_{mj}(t)] = (-i \omega_{mj} - \gamma_{mj}) M[\tilde{a}_{mj}(t)] + \sqrt{\Gamma_{mj}} M[t_{mm}(t)]
\end{equation}
with the initial condition $M[\tilde{a}_{mj}(0)] = 0$.

In the higher-order ROA, since $M[t_{mn}(t) \tilde{a}_{m'j}(t)] \neq M[t_{mn}(t)] M[\tilde{a}_{m'j}(t)]$, we evolve the reduced representation of operator products
\[
s_{mnm'j}(t) := t_{mn}(t) \tilde{a}_{m'j}(t) = i \Gamma_{m'j}^{1/2} \lim_{K\to\infty} \sum_{k \in P^K_j} \cc{g_{km'}} s_{mnk}(t)\,.
\]
Its adjoint equals $s_{mnm'j}^\dagger(t) = t_{nm}(t) \tilde{a}_{m'j}^\dagger(t)$ and the relevant commutators are $[ s_{mnm'j}(t), \tilde{a}_{m''j'}(t) ]  = 0$ and $[ s_{mnm'j}^\dagger(t), \tilde{a}_{m''j'}(t) ] $ $= - t_{nm}(t) \delta_{m'm''} \delta_{jj'}$.

Evolution equation of the system~\eqref{tmn-diffintegr} in Lorentzian bath has the form
\[
\begin{split}
&\frac{d}{dt} t_{mn}(t) = i \sum_{m'=1}^N (V_{m'm} t_{m'n}(t) - V_{nm'} t_{mm'}(t)) 
\\&\ + \sum_j \sqrt{\Gamma_{mj}} \left[ s_{mnmj}(t) - s_{nmmj}^\dagger(t) \right] + \sum_j \sqrt{\Gamma_{nj}} \left[ s_{nmnj}^\dagger(t) - s_{mnnj}(t) \right].
\end{split}
\]
Hence, the operators $s_{mnm'j}(t)$ themselves follow the evolution equation
\[
\begin{split}
&\frac{d}{dt} s_{mnm'j}(t) = i \sum_{m''=1}^N (V_{m''m} s_{m''nm'j}(t) - V_{nm''} s_{mm''m'j}(t))
\\&\ + \sum_{j'} \sqrt{\Gamma_{mj}} \left[ s_{mnmj'}(t) - s_{nmmj'}^\dagger(t) \right] \tilde{a}_{m'j}(t)
\\&\ + \sum_{j'} \sqrt{\Gamma_{nj}} \left[ s_{nmnj'}^\dagger(t) - s_{mnnj'}(t) \right] \tilde{a}_{m'j}(t) + \sqrt{\Gamma_{m'j}} \delta_{nm'} t_{mn}(t)
\\&\ + t_{mn}(t) \lim_{K\to\infty} \sum_{k \in P^K_j} \frac{\cc{g_{km'}} \omega_k e^{-i\omega_k t}}{\sqrt{\Gamma_{m'j}}} a_k(0)
 - (i \omega_{m'j} + \gamma_{m'j}) s_{mnm'j}(t)
\end{split}
\]
with the initial condition $s_{mnm'j}(0) = \frac{i}{\sqrt{\Gamma_{m'j}}} t_{mn}(0) \lim_{K\to\infty} \sum_{k \in P^K_j} \cc{g_{km'}} a_k(0)$.

Evolution equations for the reduced repretentations of the above system and interaction operators, respectively, are
\begin{equation}
\begin{split}
\label{tmnLor-reduced}
&\frac{d}{dt} M[t_{mn}(t)] = i \sum_{m'=1}^N (V_{m'm} M[t_{m'n}(t)] - V_{nm'} M[t_{mm'}(t)])
 \\&\qquad + \sum_j \sqrt{\Gamma_{mj}} \left[ M[s_{mnmj}(t)] - M[s_{nmmj}^\dagger(t)] \right] 
 \\&\qquad + \sum_j \sqrt{\Gamma_{nj}} \left[ M[s_{nmnj}^\dagger(t)] - M[s_{mnnj}(t)] \right]
\end{split}
\end{equation}
and
\begin{equation}
\begin{split}
\label{dsmnmj-dt-reduced}
&\frac{d}{dt} M[s_{mnm'j}(t)] = i \sum_{m''=1}^N (V_{m''m} M[s_{m''nm'j}(t)] - V_{nm''} M[s_{mm''m'j}(t)])
 \\&\qquad + \frac{1}{2} \sum_{j'} \sqrt{\Gamma_{mj'}} (M[s_{mnmj'}(t)] - M[s_{nmmj'}^\dagger(t)]) M[ \tilde{a}_{m'j}(t)]
 \\&\qquad + \frac{1}{2} M[s_{mnm'j}(t)] \sum_{j'} (\sqrt{\Gamma_{mj'}} M[ \tilde{a}_{mj'}(t)] - \sqrt{\Gamma_{nj'}} M[\tilde{a}_{nj'}(t)])
 \\&\qquad + \frac{1}{2}\sum_{j'} \sqrt{\Gamma_{nj'}} (M[s_{nmnj'}^\dagger(t)] - M[s_{mnnj'}(t)]) M[ \tilde{a}_{m'j}(t)]
 \\&\qquad - \frac{1}{2}\sum_{j'} ( \sqrt{\Gamma_{mj'}} M[\tilde{a}_{mj'}^\dagger(t)] - \sqrt{\Gamma_{nj'}} M[\tilde{a}_{nj'}^\dagger(t)]) M[s_{mnm'j}(t)]
 \\&\qquad - (i \omega_{m'j} + \gamma_{m'j}) M[s_{mnm'j}(t)] + \sqrt{\Gamma_{m'j}} \delta_{nm'} M[t_{mn}(t)]
\end{split}
\end{equation}
with initial condition $M[s_{mnm'j}(0)] = 0$. We use the same operator ordering scheme as in Sec.~\ref{sec:independent-baths}

Higher-order Lorentzian ROA employs \eqref{atilde-reduced}, \eqref{tmnLor-reduced} and~\eqref{dsmnmj-dt-reduced}. Its lower-order realisation describes the bath evolution with \eqref{atilde-reduced}, while for the system we represent $M[s_{mnm'j}(t)]$ as $\frac{1}{2} \{ M[t_{mn}(t)],M[\tilde{a}_{m'j}(t)]\}$, obtaining
\[
\begin{split}
&\frac{d}{dt} M[t_{mn}(t)] = i \sum_{m'=1}^N (V_{m'm} M[t_{m'n}(t)] - V_{nm'} M[t_{mm'}(t)])
 \\&\qquad+ \frac{1}{2} \bigl\{ M[t_{mn}(t)], \sum_j \bigl[ \sqrt{\Gamma_{mj}} M[\tilde{a}_{mj}(t)] 
 - \sqrt{\Gamma_{nj}} M[\tilde{a}_{nj}(t)] - \mathrm{h.c.} \bigr] \bigr\} \,,
\end{split}
\]
where the symmetrisation of the matrix product is justified by the same arguments as in Sec.~\ref{sec:genver}

The Lorentzian methods use much lower number of bath and interaction operators than the general ones, because they model the bath excitations as collective pseudomodes. Furthermore, thanks to the analytical integration of the spectral density, they automatically include the tails of the spectral density, which are cut off by the discrete method.

\section{Numerical examples and comparison with other methods}
\label{sec:examples}

In this section we demonstrate ROA by applying it to a molecular aggregate interacting with a non-Markovian quantum bath, and compare the obtained results with two other techniques: the pseudomode method\,\cite{pseudomode,pseudomode2} (PM) and the non-Markovian quantum state diffusion~\cite{strunz:1997,strunz} within an approximation called the zero-order functional expansion~\cite{nmqsd} (QSD).

The PM replaces each Lorentzian peak in the spectral density by a pseudomode with a complex frequency, and models the dynamics of the original system and bath by simulating exactly, in the Schr\"{o}dinger picture, the system interacting with this pseudomode bath. As the reduced density matrix of the system $\rho_s(t)$ obtained in this way is exact, we use the PM method as our reference. The downside of the PM method is that, since it involves an exact simulation of a quantum many-body system, its computational requirements increase exponentially with the number of bath spectral density peaks. Quite differently, QSD uses a Monte-Carlo simulation to calculate $\rho_s(t)$, facilitating the slow growth of the computational requirements with the system size. However, a numerically feasible realisation of the non-Markovian quantum state diffusion requires a further approximation such as the abstract zero-order functional expansion (ZOFE)~\cite{nmqsd} that we compare our results against. (In fact, we are not aware of any other practical, generally applicable implementation of QSD.) Furthermore, unlike our method, QSD does not allow for direct control of the total energy.

We use the above methods to model an exciton delocalised on a linear chain composed of $N$ sites coupled by the nearest-neighbour potential, $V_{mn} = J(\delta_{m,n+1} + \delta_{m,n-1})$, with $J = -1$. Each site interacts either with a single mode (Sec.~\ref{sec:example-single-mode}) or a simple zero-temperature quantum bath with a unimodal Lorentzian spectral density (Sec.~\ref{sec:example-lorentzian}).

\subsection{Single mode}
\label{sec:example-single-mode}

A simple case of a short chain, in which each site interacts with a single harmonic oscillator mode (an infinite memory bath), provides an exact reference result. We use $N=2$, $\omega_k = 4$ and $g_{km} = \sqrt{0.8}$ (intermediate system-bath coupling strength). We simulate the reduced density matrix of the system initially in the state $\Psi_s = [1, 0]^T$, and compare the probabilities of finding the system in this state at later times, i.e.~$(\rho_s(t))_{11}$.
\begin{figure}[!ht]
\centering
\includegraphics[width=0.6\columnwidth]{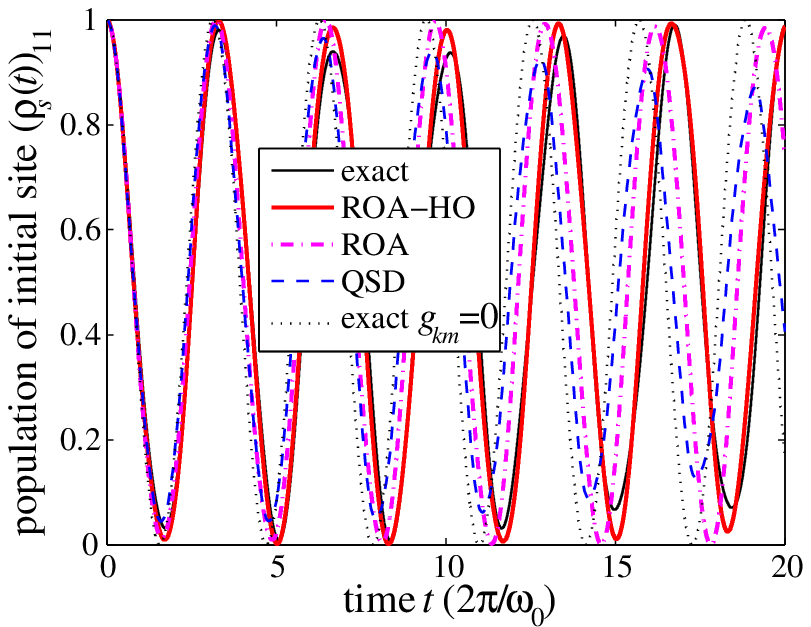}%
\caption{Dimer coupled to a single-mode bath, compared to the case of independent evolution (no system-bath coupling).}
\label{fig:single-mode}
\end{figure}

Figure\,\ref{fig:single-mode} demonstrates the results of all three approaches. The PM method applied to this case becomes simply the numerical integration of the exact Hamiltonian~\eqref{Hamiltonian}. As compared to the non-interacting solution, the probability oscillations with frequency $2|J|$ (absolute difference of energy levels of the system Hamiltonian) are modified by a phase shift and the addition of the component with frequency $\omega_k$, both resulting from the system-bath coupling. The higher-order ROA result matches the phase of the exact solution and, although it underestimates the modulation of the probability oscillations, is overall most consistent with it. On the other hand, both lower-order ROA and QSD results fail to reproduce the phase shift and oscillation modulation, with the latter method deviating the most from the exact calculation. The above comparison shows that the higher-order ROA method captures the long bath memory better than the other approaches considered.

\subsection{Lorentzian bath}
\label{sec:example-lorentzian}

In order to test ROA in the case of a finite memory bath, we consider a linear chain of $N = 3$ sites coupled to a continuous, Lorentzian bath. The bath spectral density is given by the function $J(\omega) = \Gamma \gamma \pi^{-1} / ((\omega - \omega_0)^2 + \gamma^2)$ (we set $\omega_0 = 1$). We simulate four different regimes of system-bath coupling strength and bath memory length realised by the following $\gamma$ and $\Gamma$:
\begin{center}
\begin{tabular}{l|c|c}
bath~(Fig.\,\ref{fig:baths}) &\ \ $\gamma$\ \ &\ \ $\Gamma$\ \ \\
\hline
A)\ weak narrow & 0.1 & 0.3 \\
B)\ strong narrow & 0.1 & 1 \\
C)\ weak wide & 0.5 & 0.3 \\
D)\ strong wide & 0.5 & 1 \\
\end{tabular}
\end{center}
The ``wide'' Lorentzian peaks correspond to a fast decreasing bath correlation function, while the ``narrow'' ones indicate long correlation times. The coupling strength, ``strong'' or ``weak'', determines the decoherence rate.
\begin{figure}[!ht]%
\centering
\includegraphics[width=0.54\columnwidth]{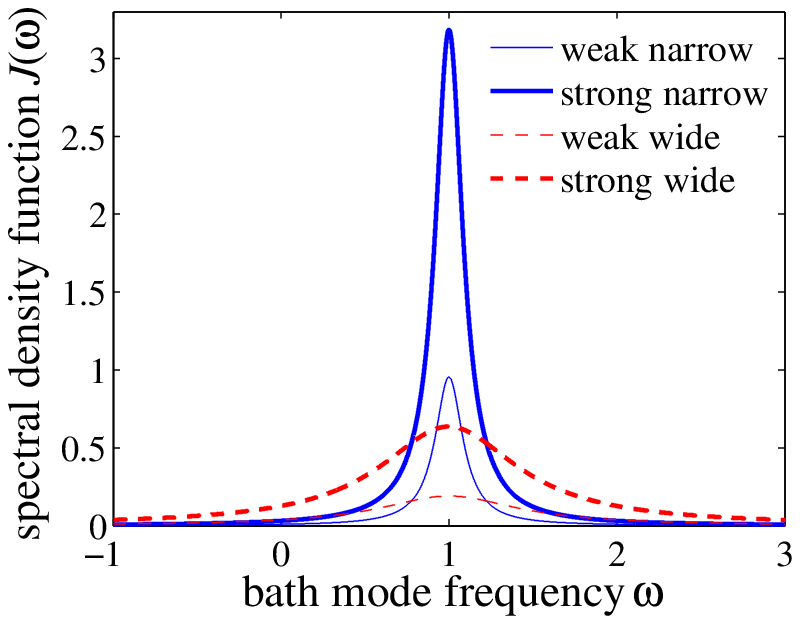}%
\caption{Unimodal Lorentzian spectral densities used in simulations.}%
\label{fig:baths}%
\end{figure}

\begin{figure}[!ht]%
\centering
\includegraphics[width=0.665\columnwidth]{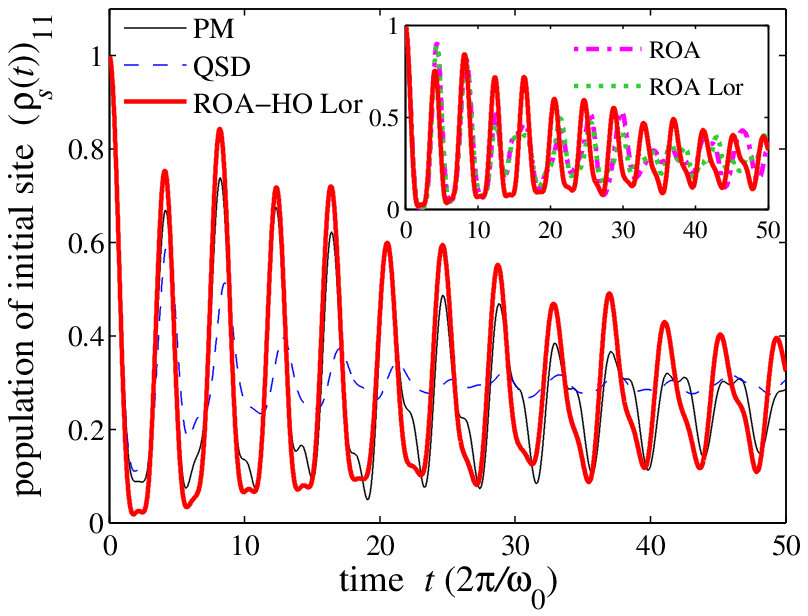}%
\caption{Comparison of ROA, PM and QSD methods for bath A. The high-order Lorentzian ROA properly reproduced the amplitude and the phase of the probability density oscillations (the inset shows that it is superior to its low order variants). The QSD method dampens the oscillations and does not reconstruct their phase.}%
\label{fig:weaknarrow}%
\end{figure}

\begin{figure}[!ht]%
\centering
\includegraphics[width=0.665\columnwidth]{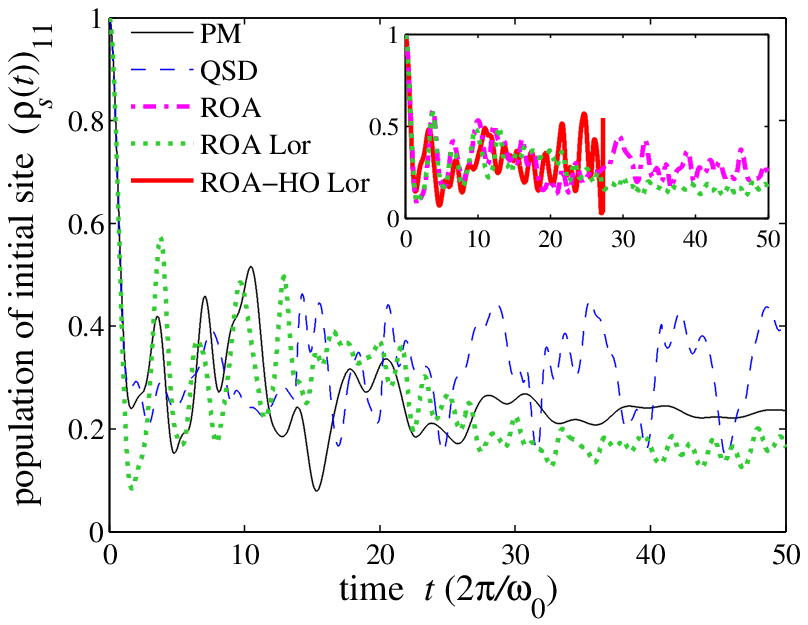}%
\caption{Comparison of ROA, PM and QSD methods for bath B. Although ROA and QSD do not reconstruct satisfactorily the PM results, the low-order Lorentzian ROA properly describes how the amplitude of fluctuations of the probability density changes in time. The inset presents a comparison of the three variants of ROA method; the high-order Lorentzian ROA diverges.}%
\label{fig:strongnarrow}%
\end{figure}

\begin{figure}[!ht]%
\centering
\includegraphics[width=0.665\columnwidth]{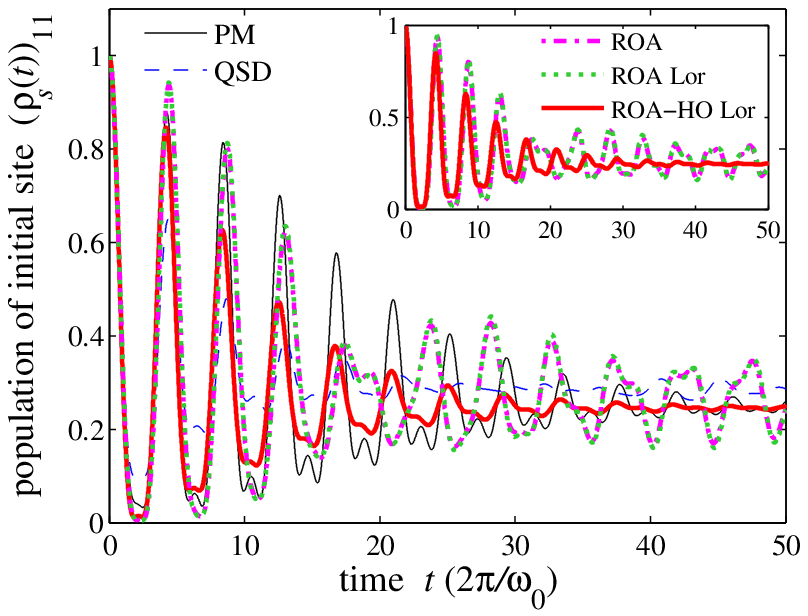}%
\caption{Comparison of ROA, PM and QSD methods for bath C. The high-order Lorentzian ROA precisely reconstructs the phase of the probability density oscillations, while its low-order variants reconstruct the amplitude. The QSD method does not describe correctly the amplitude of the oscillations and loses the phase. The inset presents the comparison of the three ROA methods.}%
\label{fig:weakwide}%
\end{figure}

\begin{figure}[!ht]%
\centering
\includegraphics[width=0.665\columnwidth]{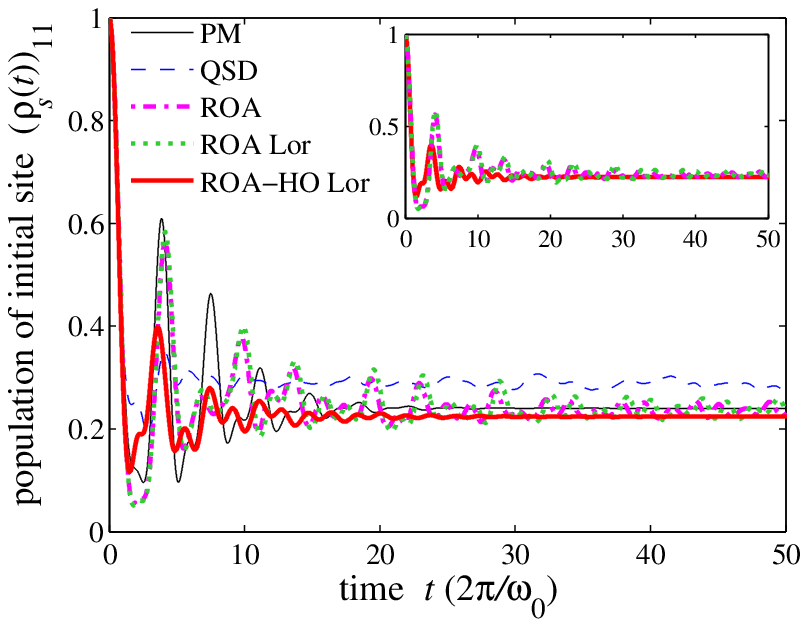}%
\caption{Comparison of ROA, PM and QSD methods for bath D. The low-order ROA methods correctly describe the amplitude and phase of the probability density oscillations at shorter times and, together with the high-order Lorentzian ROA (which diminishes the oscillation amplitudes), stabilise at the correct level. The QSD strongly diminishes the oscillations and fails to predict the correct steady state.}%
\label{fig:strongwide}%
\end{figure}

We simulate the reduced density matrix of the system, initially in the state $\Psi_s = [1, 0, 0]^T$, and compare the probabilities of finding the system in this state at later times, i.e.~$(\rho_s(t))_{11}$. We use three variants of the ROA method: low-order ROA (Sec.~\ref{sec:genver}), as well as low and high-order Lorentzian ROA (Sec.~\ref{sec:Lorentzian-spectral-densities}); the high-order ROA has been skipped as being the least efficient in this case. The results are plotted in Figs~\ref{fig:weaknarrow}--\ref{fig:strongwide}. Comparison with the exact PM method shows that the best results are obtained for the Lorentzian variants of the method, which take into account the whole range of the Lorentzian spectral density by analytical integration. At the same time QSD has the tendency to converge too rapidly to a steady state solution, as compared to the exact PM method. The figure captions contain a detailed analysis of the results.

\begin{figure}[!ht]%
\centering
\includegraphics[width=0.92\columnwidth]{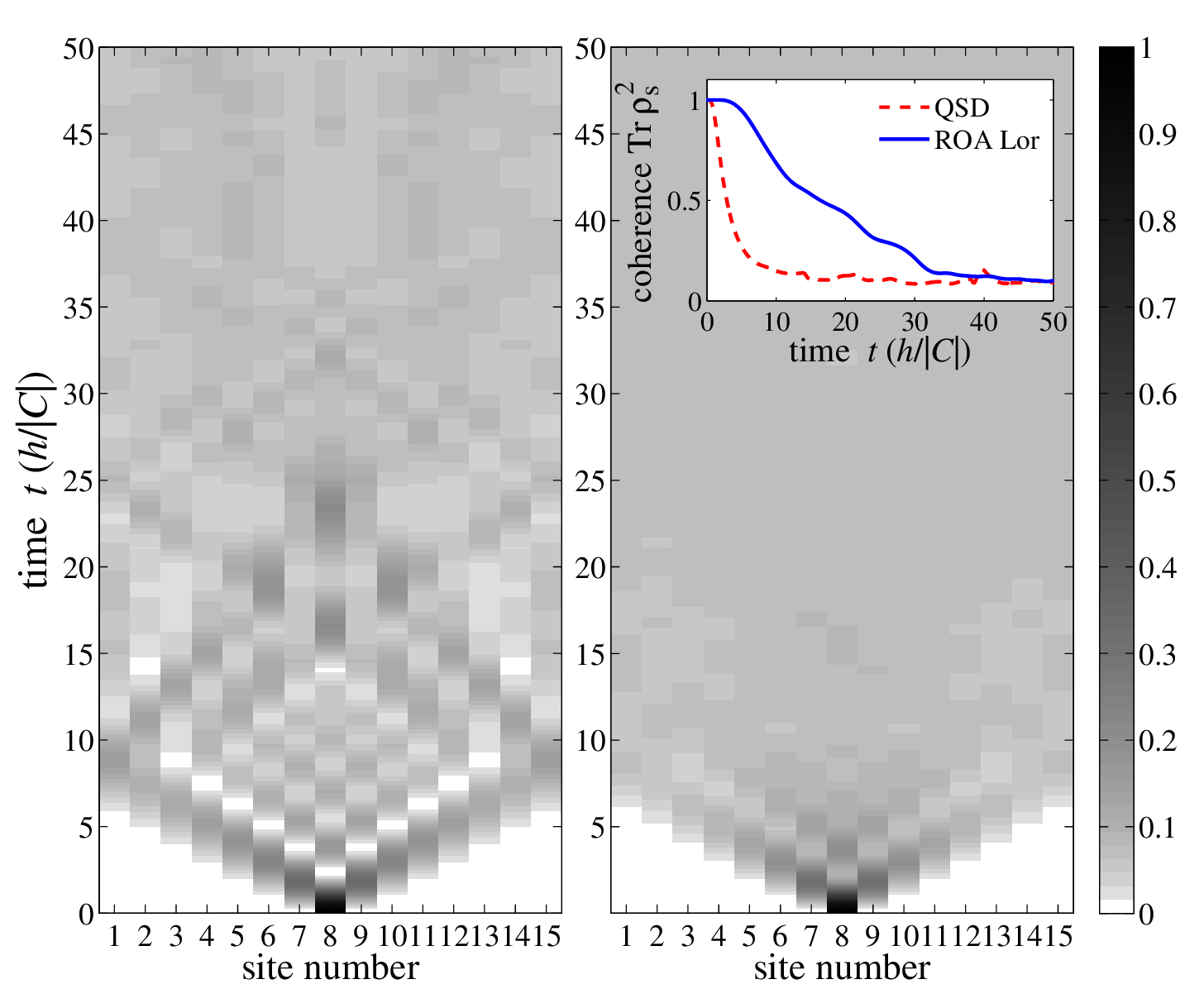}%
\caption{Comparison of the results of the ROA Lorentzian method with the QSD method for a ring aggregate of 15 sites (initially only the site 8 is excited) in multimode Lorentzian bath (see Ref.\,\cite{metoda5} for the description of the spectral density and Fig.\,3 therein for the simulation parameters used in the plot). Inset: coherence defined as ${\tr} \rho_s^2$ calculated using both methods.}%
\label{fig:transport}%
\end{figure}

For further comparison with the QSD method we calculate the transfer of the excitation on a ring aggregate in multimode Lorentzian bath (see Ref.\,\cite{metoda5}) in Fig.\,\ref{fig:transport}. The results obtained using the low-order Lorentzian ROA method are characterised by a much slower decoherence rate, as shown in the inset. This suggests that quantum coherence effects play a larger role in the dynamics of excitons in open systems than predicted in Ref.\,\cite{metoda5}. 

We attribute the observed deficiencies of the QSD method to the following factors:\\
--\ QSD treats each path of the Monte-Carlo simulation independently and thus neglects the correlation (relative phase) between the components of the joint system-bath state corresponding to different bath basis states. This approximation artificially increases the amount of decoherence in the system.\\
--\ The ZOFE approximation~\cite{nmqsd,metoda5}, required to make the method numerically feasible, treats the creation and annihilation operators of bath modes in an asymmetrical manner: while the creation operators are represented exactly and depend linearly on stochastic Gaussian noise, the annihilation operators are replaced by noise-independent operators $\bar{D}^{(m)}(t)$ acting on the system Hilbert space (see (7) in Ref.\,\cite{metoda5} and Sections~IV and~V in Ref.\,\cite{nmqsd}). Since the Gaussian noise can be arbitrarily large, this asymmetry of representation leads to an imbalance between the creation and annihilation of bath mode excitations. This artificially decreases the bath memory, as the information about the system state sent to the bath via bath mode excitations cannot ``return'' to the system when these excitations are annihilated. Similarly, since the noise---via its autocorrelation function---carries the information about the bath frequency $\omega_0$, the lack of dependence of $\bar{D}^{(m)}(t)$ operators on it and the imbalance of the creation and annihilation of bath mode excitations lead to the observed loss of phase in the occupation probability oscillations.\\
--\ From the derivation of the ZOFE operators it follows (see\,\cite{nmqsd} and (14) in Ref.\,\cite{strunz:1997}) that the ZOFE operators $\bar{D}^{(m)}(t)$ should commute with the Lindblad operators $-t_{mm}$ (in general, they should commute with every operator acting on the system Hilbert space $\Hs$ only, but this requirement would be clearly impossible to satisfy). As the evolution equation used to calculate $ \bar{D}^{(m)}(t) $ depends on the arbitrary system Hamiltonian $H_s$, one cannot hope to satisfy the condition $[-t_{mm}, \bar{D}^{(m)}(t) ] = 0$ for $t > 0$. We have verified numerically that this is indeed the case and that $[-t_{mm}, \bar{D}^{(m)}(t) ]$ increasingly deviates from zero in a QSD simulation using the ZOFE approximation. Physically, this means that the interaction between the bath and the system becomes stronger, further increasing the decohorence of the latter.

Comparing the computational efficiency of ROA and QSD methods (where we have averaged the transfer over 1000 realizations of the stochastic noise~\cite{metoda5}), for the considered system and our well-optimized C++ implementations~\cite{github} our method is more than 10 times faster on a modern desktop PC.

\section{Summary}
\label{sec:summary}

The presented Reduced Operator Approximation is a simple, physically transparent and computationally efficient method of modelling open quantum systems. It employs the Heisenberg picture of the quantum dynamics, which allows us to focus on the system degrees of freedom and capture the decoherence effects due to its coupling with the bath, while conserving the total energy. We have described different variants of the method: the low-order ROA, the high-order ROA (including the interaction operators) and their versions for Lorentzian baths. Although here it has been derived for the simplest case of linear coupling between the system and the bath, it can be easily extended to higher-order couplings. Its another advantage over methods which do not use the Heisenberg picture (such as those used in our comparisons) is that a single simulation of the reduced system operators can be used to generate reduced density matrices for an arbitrary choice of initial system state. The comparison of the ROA results for different systems (coupled to different baths with different strengths) with popular methods for modelling open quantum systems (PM and QSD) favours the ROA approach, while the efficiency of the method (especially the low-order case) is much higher than in the case of two other approaches. Furthermore, the ROA simulations of the exciton transfer on a ring aggregate suggest that quantum coherence effects persist much longer than previously thought and may play an important role in the dynamics of open quantum systems.


\section*{Appendix A: Lower-order operator representation products}
\label{sec:symmetrisation}
\def\theequation{A.\arabic{equation}}

Let us consider a system-bath Hamiltonian with $N$ system basis states and one bath mode per basis state ($K = N$), coupled only to this state ($(\mat{g}_k)_{mn} = g_k \delta_{m,k}\delta_{n,k}$). To further simplify the problem, we assume that the basis states are degenerate and have zero energies when isolated from the bath ($\mat{V} = 0$). In this case, one can solve analytically the Heisenberg equations of motion for the bath operators and radically simplify the corresponding equations for the system operators. It turns out that even in this simple case, the reduced operator matrix $M[t_{mn}(t)]$ does not commute with $M[a_k(t)]$. We will exploit this to determine what is the best way of approximating $M[t_{mn}(t) a_k(t)]$ by products $M[t_{mn}(t)]M[a_k(t)]$ and $M[a_k(t)]M[t_{mn}(t)]$, providing a justification for the choice of $\theta_l = 1/2$ in Sec.~\ref{sec:genver} The analysis we perform here is similar in spirit to the one performed when analysing the stability of discretisation schemes for partial differential equations, where a detailed analysis of a very simple problem yields important insights into the behaviour of more complex ones. We would like to stress that any numerically feasible method of modelling physical systems must be analysed from two points of view: physical foundations and numerical stability. Hence, the approximations one necessarily has to make may be motivated not only by physical principles, but also by numerical analysis.

Substituting our assumptions into \eqref{tmn-dyn-idx}, we obtain $\frac{d}{dt} t_{mm}(t) = 0$, leading to $t_{mm}(t) = t_{mm}(0)$ (our method recovers this result). Hence, we can solve \eqref{ak-dyn} explicitly, obtaining
\[
a_m(t) = e^{-i\omega_m t} \left( a_m(0) - \frac{g_m}{\omega_m} (e^{i\omega_m t} - 1) t_{mm}(t) \right)
\]
and, in the reduced representation,
\[
M[a_m(t)] = - \frac{g_m}{\omega_m} (1 - e^{-i\omega_m t}) M[t_{mm}(t)]\,.
\]

Let us assume that at time $t$, the relation $M[t_{mn}(t)] M[t_{m'n'}(t)] =$\linebreak $\delta_{nm'} M[t_{mn'}(t)]$ is preserved. Hence,
\[
\begin{split}
&M[t_{mn}(t)] M[a_k(t)] = - \delta_{nk} \frac{g_n}{\omega_n} (1 - e^{-i\omega_n t}) M[t_{mn}(t)]\,,
\\& M[a_k(t)] M[t_{mn}(t)] = - \delta_{mk} \frac{g_m}{\omega_m} (1 - e^{-i\omega_m t}) M[t_{mn}(t)]\,.
\end{split}
\]
Applying these results to \eqref{tmn-dyn-MFA}, we obtain
\[
\frac{d}{dt} M[t_{mn}(t)] = i (1 - \delta_{mn}) [ \theta_l ( \cc{\alpha_m(t)} - \alpha_n(t) )
+ (1-\theta_l )(\alpha_m(t) - \cc{\alpha_n(t)}) ] M[t_{mn}(t)]\,,
\]
where $\alpha_m(t) := - \frac{|g_m|^2}{\omega_m} (1 - e^{-i\omega_m t})$.

The Heisenberg equations of motions preserve the identity $t_{mn}(t) =$\linebreak $t_{mn'}(t) t_{n'n}(t)$. Given the fact that we use the products of $M[t_{mn}(t)]$ matrices in calculating the reduced density matrix $\rho_s(t)$ (see Sec.~\ref{sec:reduced-system-density-matrix}), we would like the equations for motion for $M[t_{mn}(t)]$ to preserve the identity $M[t_{mn}(t)] = M[t_{mn'}(t)] M[t_{n'n}(t)]$ too. One can easily verify that for almost all choices of $m$, $n$ and $n'$ this is the case for any value of $\theta_l \in [0,1]$. However, in the case of $m \neq n'$ and $n' \neq n$, we have
\[
\begin{split}
&\frac{d}{dt} M[t_{mn'}(t)] M[t_{n'n}(t)] = i [ \theta_l ( \cc{\alpha_m(t)} - \alpha_{n'}(t) ) + i [ \theta_l ( \cc{\alpha_{n'}(t)} - \alpha_{n}(t) ) \\
&\quad+ (1-\theta_l )(\alpha_m(t) - \cc{\alpha_{n'}(t)}) ] M[t_{mn}(t)] + (1-\theta_l )(\alpha_{n'}(t) - \cc{\alpha_{n}(t)}) ] M[t_{mn}(t)] \,.
\end{split}
\]
It is only for $\theta_l = 1/2$ that the terms with $\alpha_{n'}(t)$ cancel out and the right side is equal to $\frac{d}{dt} M[t_{mn}(t)]$. Therefore, only for $\theta_l = 1/2$ evolution equations for $M[t_{mn}(t)]$ preserve product identities for $t_{mn}(t)$.

Extending the above result on the case of general system Hamiltonian and general system-bath coupling, we use $\theta_l = 1/2$ in \eqref{tmn-dyn-MFA} so that the time evolution of the product $M[t_{mn}(t)] M[t_{nn'}(t)]$ is as close as possible to the time evolution of $M[t_{mn'}(t)]$.


\end{document}